\documentstyle[preprint,aps]{revtex}
\newcommand{\lsno}{La$_{1\frac{2}{3}}$Sr$_{\frac{1}{3}}$NiO$_4$}
\begin{document}
\draft
\title{\bf  Melting of Quasi-Two-Dimensional Charge Stripes in \lsno~}
\author{S.-H.  Lee$^1$ and S-W. Cheong$^2$}
\address{$^{1}$University of Maryland, College Park, MD 20742 and National
Institute of Standards and Technology, Gaithersburg, MD 20899}
\address{$^2$Bell Laboratories, Lucent Technologies, Murray Hill,
NJ 07974}
\maketitle
\begin{abstract}
Commensurability effects for nickelates have been studied by
the first neutron experiments on \lsno.
Upon cooling,
this system undergoes three successive phase transitions associated
with quasi-two-dimensional (2D) commensurate charge and spin stripe
ordering in the NiO$_2$ planes. 
The two lower temperature phases (denoted as phase II and III)
are stripe lattice states with quasi-long-range in-plane charge
correlation. When the lattice of 2D charge stripes melts, it
goes through an intermediate glass state (phase I) 
before becoming a disordered liquid state. This glass state shows
short-range charge order without spin order,
and may be called a ``stripe glass" which resembles
the hexatic/nematic state in 2D melting.
\end{abstract}


Charge and spin ordering in real space have attracted much attention
due to their role in cuprate superconductivity\cite{zaan89,salk}.
Recent neutron scattering experiments have shown that the 
real-space segregation of holes and spins in a stripe form
is associated with the anomalous suppression of superconductivity
in the $\rm La_2CuO_4$ family compounds with $\sim \frac{1}{8}$ doping
\cite{tranq95}.
These findings 
lead to the speculation that the inelastic peaks explored earlier 
in superconducting cuprates\cite{cheong91} 
are due to dynamic fluctuation of these stripes.
Such charge and spin stripes were first observed in 
the isostructural system, $\rm La_{2-x}Sr_xNiO_{4+\delta}$,
by electron diffraction\cite{chen93}
and by neutron scattering measurements\cite{tranq94}.
Mechanisms such as a Fermi-surface nesting\cite{zaan89}, 
frustrated phase separation\cite{salk} and polaron ordering\cite{chen93}
have been suggested to be responsible for the formation of stripes.
One of the issues is whether
such order is charge and/or spin driven.
Experimental results reported so far are conflicting. 
The simultaneous incommensurate
peaks due to charge and spin orders were found
in $\rm La_2NiO_{4.125}$\cite{tranq94}. 
On the other hand, in $\rm La_{1.775}Sr_{0.225}NiO_4$ 
the incommensurate charge peaks
occur at significantly higher temperature than the spin peak,
even though the in-plane charge order shows 
a shorter correlation length
than the spin order in the compound\cite{tranq_pre}. 

Commensurability has significant effects for charge and spin
ordering because of commensurate lock-in effect\cite{zach}.
For example, suppression of superconductivity in 
$\rm La_{2-x}Ba_xCuO_4$ is found when static charge and spin ordering
occurs for x=$\frac{1}{8}$\cite{tranq95}.
La$_{2-x}$Sr$_x$NiO$_4$ compounds exhibit distinct anomalies at
$x$=$\frac{1}{3}$ and 
$\frac{1}{2}$\cite{chen93,cheong94,art}. 
So far, structural correlations for 
La$_{2-x}$Sr$_x$NiO$_4$ with
$x$=$\frac{1}{3}$ and $\frac{1}{2}$
have been investigated only by electron diffraction on polycrystalline
specimens. Such data are difficult to analyze quantitatively
and lack information on magnetic correlations.
Because of increasing
difficulty in the crystal growth of La$_{2-x}$Sr$_x$NiO$_4$  
with increasing Sr concentration for $x>0.225$, 
neutron scattering for the optimal dopings 
which probes both structural and magnetic correlations
was not previously possible.
We have succeeded in growing large crystals with $x=\frac{1}{3}$
which have enabled us to extract detailed information about
the order parameter and correlation lengths for this important
composition.
Our results ``consistently" 
show charge order as the driving force behind the phase transitions
in this compound.

We have, furthermore, discovered the intriguing 2D-melting-like
transitions in this commensurate $\frac{1}{3}$ composition.
The lattice of the 3D Fe$^{2+}$/Fe$^{3+}$ ordering in Fe$_3$O$_4$
melts at T$\approx$125K in a first-order fashion, accompanied by a hysteretic 
discontinuity in the resistivity\cite{verwey}.
Recently, it has been found that 
there exists a strongly first 
order melting transition of the charge-ordered state
of the 3D perovskite 
Pr$_{\frac{1}{2}}$Sr$_{\frac{1}{2}}$MnO$_3$\cite{tomi}.
In contrast with the first-order nature of 3D melting, transitions
in 2D melting can be continuous. An interesting aspect
of 2D melting is the presence
of an intermediate glass state (the so-called hexatic/nematic state) 
between the crystalline solid
and liquid phases\cite{nelson}. 
We discovered that the melting of quasi 2D charge stripes in \lsno~ 
resembles
the solid-hexatic/nematic-liquid transition in 2D melting.

A large boule of single-crystalline \lsno~ ($\sim$7 mm diameter and 
$\sim$80 mm long) was grown using the floating zone method with
a URN-2-ZM optical furnace. 
The seed crystal and the polycrystalline
feed rod were rotated at $\sim$30 rpm in opposite directions,
and both with molten zone were moved downward at the speed of
$\sim$5 mm/hour. 
The top of the boule (8 mm long with a weight of 3.5 g) was cut for 
neutron scattering measurements and was annealed
at 1000$^o$C for 100 hours with flowing $O_2$ gas. 
The sample mosaic spread was 0.7$^o$.
The neutron scattering experiments were performed on a thermal neutron
triple axis spectrometer, BT2, and a cold neutron triple
axis spectrometer, SPINS, at the NIST reactor. 
Fixed incident and final
neutron energies were selected to be 14.8meV
for measurements on BT2
and 5meV on SPINS
using pyrolytic graphite (PG) for the monochromator and analyzer.
We use the tetragonal notation to index diffraction features throughout
this paper. All measurements were done in the (hhl) zone of reciprocal space.

For La$_{2-x}$Sr$_x$NiO$_{4+\delta}$ with excess hole
concentration $\epsilon=x+2\delta$, 
charge order and spin order are characterized by wave vectors 
$(\epsilon,\epsilon,1)$ and 
$(\frac{1}{2}+\frac{\epsilon}{2},\frac{1}{2}+\frac{\epsilon}{2},0)$, 
respectively\cite{chen93,tranq94}. 
The spin wave vector $\frac{\epsilon}{2}$ from the antiferromagnetic
reciprocal point, $(\frac{1}{2},\frac{1}{2},0)$, being
half of $\epsilon$ of charge modulation is the signature of
charge stripes as magnetic antiphase 
domain walls\cite{chen93,tranq94,hayden92}. 
We consistently found $\epsilon$=0.333(2) for \lsno, as shown
in Fig. 1 and 4(a). $\epsilon$ was not dependent on temperature (T) 
(see Fig. 4(a)).
This contrasts with the considerable temperature dependence of $\epsilon$ 
when $\epsilon < \frac{1}{3}$\cite{tranq94,tranq_pre}
and indicates the stability of the $\epsilon$=$\frac{1}{3}$ modulation. 
Fig. 1 shows $(\frac{1}{3},\frac{1}{3},l)$ scans at various temperatures.
Even integer weak $l$ peaks are purely magnetic because they were
not detected by electron diffraction\cite{chen93}.
For odd integer $l$, charge-order scattering is dominant\cite{comm4}.
The superlattice peaks have long tails along $l$,
indicating the quasi-two-dimensionality of structural and spin correlations.
Finite repulsive Coulomb interaction will favor a
stacking of charge stripe layers in such a way as to minimize 
the Coulomb interaction.
Such a stripe pattern consistent with the superlattice peaks
is illustrated at the upper right corner. Notice that
all charge stripes in all layers are centered
between Ni and O atoms\cite{comm2}. 

In Fig. 1, the relative ratios of the peak intensities
vary in different T regimes. This led us into detailed measurements
of those superlattice peak intensities as a function of T, which revealed
surprising indications of successive phase transitions.
Fig. 2 summarizes the results (in (a), (b), and (c))
along with a bulk susceptibility measurement, $d(\chi T)/dT$ (in (d)).
The onset of the superlattice peaks - 
especially the $(\frac{1}{3}, \frac{1}{3},5)$ 
reflection - at $\sim$ 240K, shown in Fig. 2 (a), signals
the development of the localization of the holes
on a time scale, $\tau > 2\hbar/\Delta E \sim$ 5 ps set by
the energy resolution of the instrument.
Upon cooling, however, they develop not as 
an ordinary phase transition, but with an
unusually slow rate followed by a fast enhancement at $\sim$ 200K.
As T decreases further down to 10K, the peaks continue to grow.
At T lower than 40K, the $(\frac{1}{3}, \frac{1}{3},1)$ peak decreases.
The existence of the three T regions below 239K
is clearer in the plot of the ratio of the 
$(\frac{1}{3},\frac{1}{3},1)$ peak
intensity to the 
$(\frac{1}{3},\frac{1}{3},5)$ one
as depicted in Fig. 2(b).
These three distinct phases are denoted 
as phase I, II, III with decreasing T order.
Fig. 2 (c) shows 
the T-dependence of the pure spin contribution
to the $(\frac{1}{3}, \frac{1}{3},0)$ peak.
The contribution of the long tail of the $(\frac{1}{3}, \frac{1}{3},1)$ 
charge peak to the $(\frac{1}{3}, \frac{1}{3},0)$ peak
was assumed to be same as that at $(\frac{1}{3}, \frac{1}{3},0.3)$
and subtracted. The non-zero intensity below $\sim$ 190K 
demonstrates that the spin ordering occurs 
in phase II and III, but not in phase I (190K$<$T$<$239K). 
Anomalies in $d(\chi T)/dT$, shown in Fig. 1 (d), establish
that the charge/spin correlations in the three phases are static
on a time scale more than a few seconds.
No thermal hysteresis was observed in any of the measurements
suggesting the phase transitions are continuous.

Interesting aspects of the three phases were discovered in 
elastic Q-scans along the (hh0) and $(00l)$ directions.
The (hh0) scan investigates {\it longitudinal} correlation of 
translational ordering of stripes along the perpendicular direction
to the stripes in the NiO$_2$ plane.
Fig. 3 (a) shows representative in-plane (hh1) scans, which
probe in-plane structural correlations due to charge order.
Note that the peak is resolution limited in phase II and III but
in phase I it is significantly broader than the q-resolution. 
Thus, the in-plane charge correlation 
is much shorter in phase I than in phase II and III where the correlation
is quasi-long-range.
In contrast, the out-of-plane scans shown in Fig. 3(b) 
are much broader than the q-resolution for all phases. 
Quantitative values of those correlation lengths
were obtained by fitting each data
to a Lorentzian convoluted
with the instrumental resolution function. The fits are drawn as solid lines
in Fig. 3,
and $\epsilon$, in-plane charge ($\xi_a^C$)
and spin ($\xi_a^S$) correlation lengths, 
and out-of-plane charge ($\xi_c^C$) and spin ($\xi_c^S$) 
correlation lengths are plotted in Fig. 4.
For all T$<$239K, $\xi_c^C$ and $\xi_c^S$ are shorter than 2.5c, indicating
that the quasi-two dimensionality of both charge and spin correlations
remains intact in all three phases.
A dramatic difference between the low-T phases (II and III)
and the higher T phase (I) is manifested in the in-plane charge correlation.
As illustrated in Fig. 4 (b),
the in-plane charge correlation 
is quasi-long-range in phase II and III 
($\xi_a^C \sim 350 \AA$)
, which is consistent with the previous electron 
diffraction result\cite{chen93}.
However, in phase I, $\xi_a^C$ drastically shortens to $\sim 80\AA$, which
is about 5 times the characteristic wavelength of
charge modulation ($16.23\AA$).
There seems to be only subtle differences
between phase II and III, both of which exhibit quasi-long-range
in-plane charge correlation. The transition is probably a 
subtle change of structural distortion associated with the hole lattice. 
$\xi_a^C$ is about three times longer than $\xi_a^S$,
indicating that charge ordering is the driving force in this compound
and spin ordering is a parasitic effect.
Phase I shows short-range charge order
in the absence of spin order, which
is drastically different from the other two phases.
This intermediate glassy state with short-range
charge stripe correlation in between solid states (phase II and III)
and a liquid state (a disordered high-T state) may be called
a ``stripe glass".


As in $d(\chi T)/dT$, specific heat ($C$) exhibits a distinct anomaly
at the liquid-glass transition at 239K and a small kink
at the glass-solid transition at 190K\cite{art}, which resembles 
that seen in a liquid crystal\cite{huang}.
This indicates that the majority of excess holes localizes into
domain walls below 239K even though the longitudinal correlation
of the holes remains short-range.
As a result, the entropy removed through the glass-solid transition at 190K
is reduced, together with the anomaly.
Preliminary inelastic neutron scattering measurements\cite{shl}
show that low energy charge/spin fluctuations below 3meV are strongest 
at $\sim$200K at which the correlations become quasi long-range.
Specific heat is sensitive to fluctuations with characteristic energy of
order $\rm k_BT$. Hence, the distinct specific heat anomaly 
and the weakness of neutron
scattering cross section below 3 meV at 239K ($\sim$ 20meV)
imply that the characteristic energy of charge fluctuations in the stripe
glass state is much higher than 3 meV.

The short-range translational ordering and the specific heat behavior
are reminiscent of the 2D glass in 2D melting.
Missing in this connection is the orientational order parameter.
The 2D glass in 2D melting
is characterized by short-range positional
order and quasi-long-range orientational order,
and has been tested for various real systems 
such as liquid crystals\cite{huang},
vortex states of layered cuprate superconductors\cite{grier},
and doped charge-density-wave compounds\cite{dai}. 
The nematic phase in the liquid
crystal has the strongest similarity with the stripe glass
in the sense that both of them have two-fold symmetry\cite{liquidc}.
The orientational order parameter in such systems 
is an average orientation of stripes/molecules.
The difference is, however, that the stripes can have various
lengths, whereas molecules in nematic phase have the same length.
One might think that the orientational order parameter for the stripes
can be measured by {\it transverse} scans. 
However, when the stripes are of various lengths
even though they are aligned along a common axis, 
the broadening of transverse scan peaks
due to short stripes overshadows the characteristic width 
due to the orientational order and makes it nearly impossible
to extract orientational order parameter from the scans.
Nonetheless, the 2D nature of charge order in \lsno~ 
and the presence of an intermediate state with
short-range ordering 
show the strong similarity of the stripe glass with
the 2D glass state.
It remains in question as to
whether or not such an intermediate glass state in $\rm La_{2-x}Sr_xNiO_4$
system exists only when x=$\frac{1}{3}$. 
More experimental studies on x$>\frac{1}{3}$ compounds, especially for 
x=$\frac{1}{2}$ where bulk measurements show distinct anomalies,
as well as theoretical studies, are needed to understand the commensurability
effect in the system.

The impurity potential of Sr ions may play an important
role in the melting transition of the charge stripes.
Sr ions are distributed randomly on the La sites in the La-O layers
and can provide pinning potentials for the charge ordering with stripe
modulation. 
These point-defect-like Sr ions can be pinning centers
for the translational ordering of the charge stripes, 
but are naturally
inefficient in pinning the orientational ordering of charge stripes. 
This pinning effect of Sr dopants may be responsible for a stripe glass
at intermediate temperatures, 190K$<$T$<$239K, where
the mechanism for producing stripe phase does not dominate
the impurity potential and thermal fluctuations.
In that phase, most of the holes localize to form strings (possibly
maintaining orientational ordering), but the lengths of the strings
are short.
Upon further cooling, these short strings of holes line up straight to form
ordered stripes. Due to the frozen defects, there would be defect regions
where two stripes merge into one stripe to form a ``double-fork" stripe
defect. Such defects would create topological spin frustration,
while only producing a small local perturbation of the charge stripe order.
This type of stripe defects may be responsible for the shorter
correlation of spin than that of charge in the phases II and III.

In summary, 
$\rm La_{2-x}Sr_xNiO_4$ exhibits, at x=$\frac{1}{3}$, unique behaviors
which are typical for melting in two dimensions. 
Among three successive transitions upon cooling,
the first two transitions are the continuous
transitions of the liquid-glass-solid states of quasi-2D hole ordering.
We also presented ``consistent" evidences,
namely order parameter and correlation length measurements,
showing that charge ordering is the driving force in this material.

We have benefitted greatly from discussions with P.A. Anderson,
C.L. Broholm, C.H. Chen, B.I. Halperin, Y. Ijiri, P. Littlewood, 
and O. Zachar.
Work at SPINS is based upon activities supported by the National
Science Foundation under Agreement No. DMR-9423101.

\begin{figure}
\caption{
Neutron diffraction data along $(\frac{1}{3},\frac{1}{3},l)$
at a series of temperatures. Data were taken
on BT2 with collimations of 60$^{'}$-40$^{'}$-40$^{'}$-80$^{'}$.
Energy resolution was 1.4(1)meV.
Typical error was 5\% of the intensity.
Backgrounds were determined by the measurement at 300K and were subtracted.
At the upper right corner is a stripe pattern for $\epsilon=\frac{1}{3}$
drawn.
Filled and open circles indicate Ni sites in one NiO$_2$ plane and
in an adjacent plane, respectively.
Square lines, solid and dotted, are the square lattices of Ni atoms.
Diagonal lines, solid and dotted, indicate positions of charge stripes
in each plane.
}
\end{figure}

\begin{figure}
\caption{
Elastic neutron scattering intensities of several superlattice
peaks and bulk susceptibility data.
The energy resolution for the neutron scattering measurements on SPINS
was 0.28(2)meV with collimations of open-40$^{'}$-40$^{'}$-open.
(a) Temperature dependences of the superlattice 
peak intensities at Q=$(\frac{1}{3},\frac{1}{3},1)$, 
$(\frac{1}{3},\frac{1}{3},3)$,
and $(\frac{1}{3},\frac{1}{3},5)$. The data are in arbitrary units.
(b) Ratio of the 
$(\frac{1}{3},\frac{1}{3},1)$ 
peak intensity to $(\frac{1}{3},\frac{1}{3},5)$ peak. 
(c) Temperature dependence of the magnetic superlattice peak intensity.
Units are the same as those in (a).
(d) d($\chi\cdot \rm T$)/dT in emu/mole. The line is a guide to the eye.
}
\end{figure}
\begin{figure}
\caption{
(a) Representative elastic (hh1) scans for phase I (220K),
phase II (60K) and phase III (10K).
(b) Representative elastic ($\frac{1}{3},\frac{1}{3},l$)
scans in each phase taken from Fig 3. 
Solid lines in (a) and (b) are explained in the text.
Q-resolutions are drawn as bars.
}
\end{figure}
\begin{figure}
\caption{
$\epsilon$ and correlation lengths versus temperature.
T-dependence of (a) $\epsilon$ obtained from optimal peak
positions of ($\frac{1}{3},\frac{1}{3},1$) peaks,
(b) in-plane correlation lengths, $\xi_a^C$ (charge) and $\xi_a^S$ (spin), 
obtained by fitting ($\frac{1}{3}+h,\frac{1}{3}+h,1$) and 
($\frac{1}{3}+h,\frac{1}{3}+h,0$) scans, respectively, and
(c) out-of-plane correlation lengths, $\xi_c^C$ (charge) and $\xi_c^S$ (spin), 
obtained by fitting ($\frac{1}{3},\frac{1}{3},1+l$) and 
($\frac{1}{3},\frac{1}{3},2+l$) scans.
}
\end{figure}


\begin{references}
\bibitem{comm4} The odd $l$ peaks are magnetic as well as structural.
However, considering the relative weakness of purely magnetic even $l$
peaks and the different behavior of the odd $l$ peaks 
in order parameter measurements
and in correlation lengths from those of the even $l$ peaks, 
we conclude that charge-order scattering is dominant in the odd $l$ peaks.
\bibitem{comm2} The possibility of an anisotropic distribution
of holes around Ni ions was discussed in Ref. \cite{zaanen}.
Here we assumed that each plane has the same stripe pattern.
If this assumption is weakened, a pattern in which stripes
in layers alternate between Ni- and O-centered positions is also possible.
\end{references}
\end{document}